\documentclass[twocolumn,secnumarabic,nobibnotes, aps, prl, superscriptaddress,floatfix]{revtex4-2}

\usepackage[normalem]{ulem} 	
\usepackage[usenames,dvipsnames]{color}
\usepackage{upgreek}
\usepackage{braket}
\usepackage{float}
\usepackage[nointegrals]{wasysym}
\usepackage{helvet}
\usepackage[hypertexnames=false]{hyperref}
\usepackage{xcolor}
\usepackage[english]{babel}
\usepackage{graphicx, import}
\usepackage{amssymb,amsfonts,amsmath} 
\DeclareMathOperator{\Tr}{Tr}

\renewcommand{\phi}{\varphi}

\renewcommand{\thefigure}{\arabic{figure}}
\renewcommand{\thetable}{\arabic{table}} 

\begin{document}

\title{Slowing down light in a qubit metamaterial }

\author{Jan David Brehm}
	\affiliation{Physikalisches Institut, Karlsruhe Institute of Technology, 76131 Karlsruhe, Germany}
\author{Richard Gebauer}
	\affiliation{Institute for Data Processing and Electronics, Karlsruhe Institute of Technology, 76131 Karlsruhe, Germany}
\author{Alexander Stehli}
	\affiliation{Physikalisches Institut, Karlsruhe Institute of Technology, 76131 Karlsruhe, Germany}			
\author{Alexander N. Poddubny}
	\affiliation{Ioffe Institute, St. Petersburg 194021, Russia}
\author{Oliver Sander}
	\affiliation{Institute for Data Processing and Electronics, Karlsruhe Institute of
		Technology, 76131 Karlsruhe, Germany}	
\author{Hannes Rotzinger}
	\affiliation{Physikalisches Institut, Karlsruhe Institute of Technology, 76131 Karlsruhe, Germany}
		\affiliation{Institute for Quantum Materials and Technologies, Karlsruhe Institute of Technology, 76021 Karlsruhe, Germany}
\author{Alexey V. Ustinov}
	\affiliation{Physikalisches Institut, Karlsruhe Institute of Technology, 76131 Karlsruhe, Germany}
	\affiliation{Institute for Quantum Materials and Technologies, Karlsruhe Institute of Technology, 76021 Karlsruhe, Germany}
	
	\affiliation{National University of Science and Technology MISiS, 119049 Moscow, Russia}
	\affiliation{Russian Quantum Center, Skolkovo, 143025 Moscow Region, Russia}

\date{\today}

\begin{abstract}
The rapid progress in quantum information processing leads to a rising demand for devices to control the propagation of electromagnetic wave pulses and to ultimately realize a universal and efficient quantum memory. While in recent years significant progress has been made to realize slow light and quantum memories with atoms at optical frequencies, superconducting circuits in the microwave domain still lack such devices. Here, we demonstrate slowing down electromagnetic waves in a superconducting metamaterial composed of eight qubits coupled to a common waveguide, forming a waveguide quantum electrodynamics system. We analyze two complementary approaches, one relying on dressed states of the Autler-Townes splitting, and the other based on a tailored dispersion profile using the qubits tunability. Our time-resolved experiments show reduced group velocities of down to a factor of about 1500 smaller than in vacuum. Depending on the method used, the speed of light can be controlled with an additional microwave tone or an effective qubit detuning. Our findings demonstrate high flexibility of superconducting circuits to realize custom band structures and open the door to microwave dispersion engineering in the quantum regime.
\end{abstract}

\maketitle 


\section{Introduction}
In the light of quantum information processing achieving control over the speed of light in artificially structured media, slowing it down and eventually stopping the light, has recently regained attention. This functionality is of vital importance for the realization of long-living quantum memories \cite{lvovsky_optical_2009} and for controlling and synchronizing the flow of information \cite{kroh_slow_2019}.
One of the most prominent techniques to realize slow light is based on electromagnetically induced transparency (EIT) - a quantum interference effect between different excitation pathways, rendering an otherwise opaque medium transparent and creating a steep dispersion profile \cite{fleischhauer_electromagnetically_2005}. Early demonstrations of slow light in EIT media involved vapors of sodium \cite{hau_light_1999-1} and rubidium atoms \cite{kash_ultraslow_1999}, followed by further experiments with cold atoms coupled to nanofibers \cite{gouraud_demonstration_2015}, and eventually by the demonstration of a highly efficient quantum memory \cite{hsiao_highly_2018}. A promising candidate beyond atoms to observe these effects are superconducting qubits in the context of cavity-free waveguide quantum electrodynamics (wQED). The observation of resonance florescence revealed that even single qubits can have strong coupling to propagating electromagnetic waves and show extinction coefficients close to unity  \cite{astafiev_resonance_2010,hoi_microwave_2013}. First demonstrations of the single-qubit Autler-Townes splitting (ATS) with ladder-type three-level systems \cite{abdumalikov_electromagnetically_2010,hoi_microwave_2013}, showcased their potential suitability for the EIT-related applications. However, it was subsequently argued that in these experiments the EIT regime was not unambiguously reached \cite{anisimov_objectively_2011}.

More recently, focus shifted towards multi-qubit systems with infinite range interactions \cite{lalumiere_input-output_2013,loo_photon-mediated_2013}, super- and subradiant polaritonic excitations \cite{brehm_waveguide_2020,loo_photon-mediated_2013,zhang_theory_2018,zhong_photon-mediated_2020}, collective ATS \cite{brehm_waveguide_2020},  topological edge states \cite{ke_radiative_2020, poshakinskiy_quantum_2020}, creation of non-classical light \cite{fang_one-dimensional_2014}, and proposals for superconducting quantum memories \cite{leung_coherent_2012}. Even though superconducting wQED systems can match the requirements of large optical depth and high coherence \cite{everett_stationary_nodate}, slow light has been so far realized only in the context of classical waveguides \cite{mirhosseini_superconducting_2018-1}. These devices are passive, without the possibility to in-situ control the speed of light and to ultimately realize a quantum memory protocol.

Here, we experimentally demonstrate a first realization of slow light in a superconducting wQED system consisting of eight locally tunable transmon qubits coupled to a one-dimensional waveguide. We engineer the required flat band structure by using qubits directly as controllable dispersive elements. First, we consider the standard case of dressed state based slow light and demonstrate that even in the case of imperfect EIT, where the physics is merely governed by the ATS, a moderate retardation of the group velocities by a factor down to 1500 compared to vacuum can be achieved. Second, we engineer a similar band structure, based on detuned collective resonances of the participating qubits. This allows for three times larger efficiencies of the slow-light medium. The demonstrated slow light effect can be used for a fixed-delay quantum memory in superconducting wQED and paves way to more general applications such as on demand storage-and-retrieval memory for quantum information processing. 

\section{Results and Discussion}
\subsection{Circuit design and properties}

\begin{figure}[t!]
	\includegraphics[width=\columnwidth]{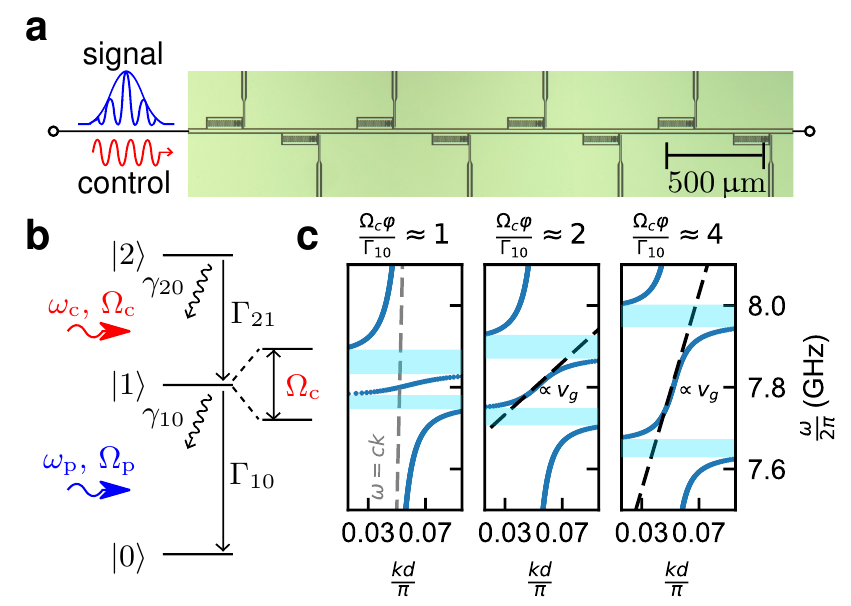}
	\caption{\textbf{a}. Microscopic image of the eight qubit metamaterial with local flux bias lines. Both signal and control tone propagate along the central waveguide. \textbf{b}. Ladder-type level structure of the employed transmon qubits and the relevant radiative relaxation ($\Gamma_{10},\ \Gamma_{21}$) and decoherence rates ($\gamma_{20},\ \gamma_{10}$). If the control tone is resonant with the $2\rightarrow 1$ transition, the first transmon level hybridizes, creating the Autler-Townes splitting. \textbf{c}. Calculated band structure of the metamaterial  in the lossless limit for different control Rabi-strengths $\Omega_\text{c}$. Light blue shaded areas indicate bandgaps. The effective group velocity $v_\text{g}$ in the center band strongly depends on $\Omega_\text{c}$.}
	\label{fig1}
\end{figure}

The qubit metamaterial presented in this work consists of eight superconducting transmon qubits \cite{koch_charge-insensitive_2007}, capacitively coupled to the mode continuum of a coplanar waveguide, see Fig.~\ref{fig1} a. Each qubit has an integrated SQUID loop and is thereby individually tunable between 3 and 8$\,$GHz by applying local magnetic flux via a flux bias line. We fulfill the metamaterial limit of sub-wavelength dimensions by choosing a dense qubit spacing of $d=400\,\upmu$m, which corresponds to a fraction of the light wavelength $\lambda$ with the phase delay of $\phi=\frac{2\pi}{\lambda}d=0.05-0.16$. As shown in former works, in such a setting the qubits obtain an infinite-range photon-mediated effective interaction, which is almost exclusively of collective dissipative nature \cite{lalumiere_input-output_2013, loo_photon-mediated_2013}. For a detailed study of the mode structure of this metamaterial, its super- and subradiant polaritonic excitations, we guide the reader to reference \cite{brehm_waveguide_2020}. Figure~\ref{fig1} b shows the first three ladder-type energy levels of the transmon qubits. If the $2\rightarrow 1$ transition is driven with a microwave control tone with the amplitude (Rabi-strength) $\Omega_\text{c}$, these levels hybridize and split into two dressed states separated by $\Omega_\text{c}$, forming the ATS \cite{autler_stark_1955}. For our experiments we use the ATS of the individual qubits to calibrate the absolute value of the control power $P_\text{c}=\frac{\Omega_\text{c}^2}{4\Gamma_{10}}\hbar\omega_\text{c}$ \cite{honigl-decrinis_two-level_2020} (see Supplementary Material). With reference to a weak probe tone and assuming a $\propto\exp(-\text{i}\omega t)$ time dependence, the reflection coefficient of a single qubit is given by \cite{abdumalikov_electromagnetically_2010}:
\begin{equation}
r=-\frac{\Gamma_{10}}{2[\gamma_{10}-\text{i}(\omega-\omega_{10})]+\frac{\Omega_c^2}{2\gamma_{20}-2\text{i}(\omega-\omega_{10}+\omega_c-\omega_{21})}}
\label{reflection}
\end{equation}
The corresponding transmission coefficient is given by $t=1+r$, referenced in this article as element of the scattering matrix $S_{21}(\omega)$.	
Here, the qubits are strongly coupled to the waveguide, ensuring a multi-mode Purcell limited average relaxation rate of $\Gamma_{10}/2\pi=12\,$MHz. The average decoherence rate of the $1\rightarrow 0$ transition is $\gamma_{10}/2\pi=(\Gamma_{10}/2+\Gamma_\text{nr})/2\pi=6.9\,$MHz, with $\Gamma_\text{nr}$ accounting for pure dephasing and radiative losses to unguided modes. The corresponding average decoherence rate of the $2\rightarrow 0$ transition is $\gamma_{20}/2\pi=6.9\,$MHz, which is here by coincidence the same value. We note that this rate is, in agreement with other experiments with superconducting qubits \cite{abdumalikov_electromagnetically_2010, hoi_demonstration_2011}, significantly higher than that of cold atoms \cite{fleischhauer_electromagnetically_2005}. The increased rate is due to the ladder-type level structure of the transmon, where it can be shown that $\gamma_{20}>\Gamma_{21}/2$ always holds \cite{yan_electromagnetically_2001}. This necessarily leads to imperfect EIT since the associated dark state obtains a finite lifetime \cite{fleischhauer_electromagnetically_2005}. The corresponding implications for the creation of slow light are discussed in the following section. The expected band structure for the metamaterial in the lossless case ($\Gamma_\text{nr}=0$, $\gamma_{20}=0$) based on Eq.~\eqref{reflection} for different values of $\Omega_\text{c}$ is shown in Fig.~\ref{fig1} c. The expected group velocity of light $v_\text{g}=\frac{d \omega}{d k } $ in the center band strongly depends on the control tone strength $\Omega_\text{c}$. In the regime of small $\Omega_\text{c}$ the slope of the band is almost flat, giving rise to slowing down the electromagnetic waves in the waveguide.

\subsection{Slow light based on dressed-state Autler-Townes splitting}

\begin{figure}[tb!]
	\includegraphics[width=\columnwidth]{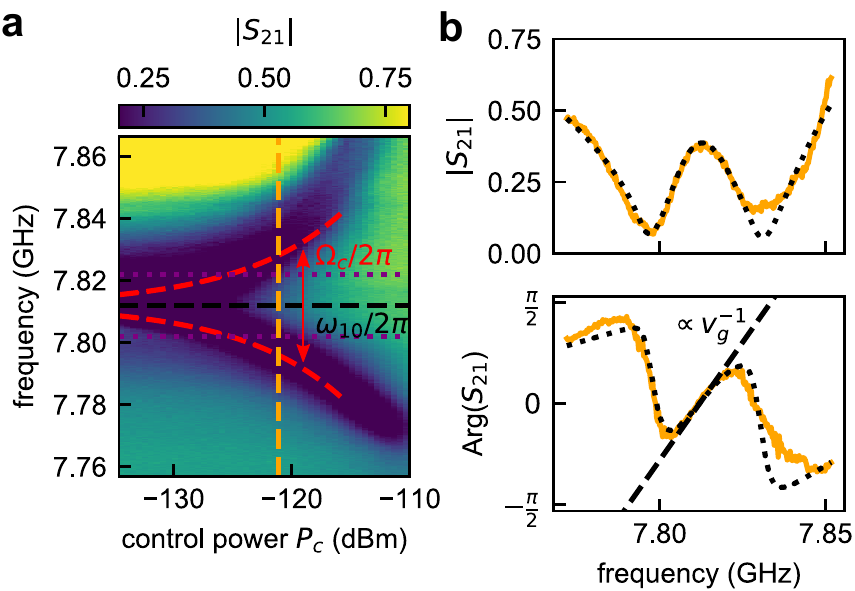}
	\caption{\textbf{a}. Collective ATS of seven resonant qubits in dependence on control tone power $P_\text{c}$.  For $P_\text{c}>-125\,$dBm, a transparency window with finite transmission emerges around the resonance frequency $\omega_\text{10}$ (black dashed line). The red dashed line is a fit to the expected splitting of $\Omega_\text{c}/2\pi$. Purple dotted lines mark a minimal bandwidth of $1/\sigma=20\,$MHz of the transparency window for time-domain experiments with $\sigma=50\,$ns pulses. \textbf{b}. Absolute value and phase of the complex transmission coefficient $S_{21}$ for $P_\text{c}=-122\,$dBm. Black dotted lines are fits to a transfer-matrix model. The effective expected group velocity $v_\text{g}$ is anti-proportional to the slope of $\text{Arg}(S_{21})$ at $\omega_\text{10}$.}
	\label{fig2}
\end{figure}

In order to create a slow-light medium, the qubits are consecutively tuned to a common resonance frequency $\omega_{10}/2\pi=7.812\,$GHz in the vicinity of the qubits' upper sweet spot. A continuous microwave control tone is applied, driving resonantly the $2\rightarrow 1$ transition at the frequency $\omega_\text{c}/2\pi=7.533\,$GHz and Rabi-strength $\Omega_\text{c}$. Figure~\ref{fig2} shows the collective ATS, for $N=7$ resonant qubits. For low control powers $P_\text{c}\rightarrow 0$, only a single bandgap of strongly suppressed transmission above $\omega_{10}$ is present, which gradually splits for higher $\Omega_\text{c}$ into two separate bandgaps, with a transparency window of finite transmission opening up in between (compare band structure calculation in Fig.~\ref{fig1} c). In this region the phase of the complex transmission coefficient Arg$(S_{21}(\omega))$ features a steep roll-off (Fig.~\ref{fig2} b), indicating low group velocities. The observed transmission coefficient is in agreement with a numerical transfer-matrix calculation \cite{brehm_waveguide_2020,asenjo-garcia_exponential_2017}, which takes into account a cable resonance caused by impedance mismatches in the cryostat wiring. Due to the aforementioned large $2\rightarrow 0$ decoherence rate $\gamma_{20}\approx\gamma_{10}$, the expected sharp EIT feature at $\omega_{10}$ with near unity transmission for small $\Omega_\text{c}$ is smeared out and transmission is reduced. The expected effective group velocity $v_\text{g}$ of a signal sent through the transparency window can be calculated from the phase gradient \cite{leung_coherent_2012, asenjo-garcia_exponential_2017}:
\begin{equation}
v_\text{g}=\left.\left[\frac{1}{(N-1)d}\frac{d\text{Arg}(S_{21}(\omega))}{d\omega}\right]^{-1}\right\vert_{\omega=\omega_{10}}
\label{finitesystemdelay}
\end{equation}

\begin{figure}[htb!]
	\includegraphics[width=\columnwidth]{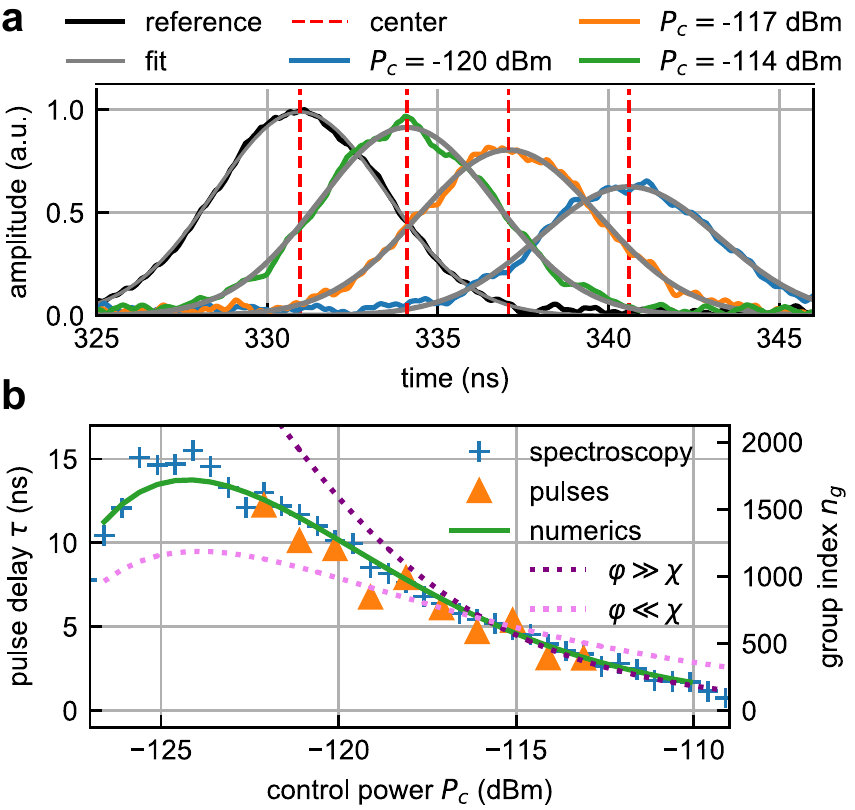}
	\caption{\textbf{a}. Envelope of the detected Gaussian pulses in the time domain for different control tone powers $P_\text{c}$ of the $N=7$ qubit ATS. The pulses are generated with a center frequency of $\omega_\text{10}=7.812\,$GHz and width $\sigma=50\,$ns, corresponding to the center of the transparency window. For better visibility the pulses are compressed by a factor of 20, where their maximum remains at the original position. \textbf{b}. Experimentally extracted delays from pulsed and spectroscopic measurements in dependence on control power $P_\text{c}$. Pulse delays of up to $12\,$ns ($15\,$ns in spectroscopy) were achieved, corresponding to the reduction of the group velocity of 1500 (1900) compared to the speed of light in vacuum.}
	\label{fig3}
\end{figure}

The traversal time $\tau=(N-1)d/v_\text{g}$ of a pulse through the medium with $N=7$ qubits inferred from the spectroscopic data of Fig.~\ref{fig2} a at $\omega=\omega_{10}$ is shown in Fig.~\ref{fig3} b. In good agreement with the numerical transfer-matrix model the expected traversal time, or conversely the effective group index $n_\text{g}$, can be tuned over a large range with the applied control power $P_\text{c}$.
In contrast to the textbook case of EIT in the limit of $\gamma_{20}\approx 0$, where $\tau\propto 1/\Omega_\text{c}^2$ diverges for small $\Omega_\text{c}$ \cite{fleischhauer_electromagnetically_2005}, we observe $\tau$ approaching a maximum of $15\,$ns ($n_\text{g}\approx 1900$) at $P_\text{c}\approx-124\,$dBm before it decreases again for even lower control tone strengths. A similar behavior was observed in room temperature vapor of $^4$He \cite{goldfarb_electromagnetically-induced_2009}. An analytical expression for the expected delay $\tau(\omega_{10})$ in the case of a resonant control tone can be found from the effective dispersion relation $(kd)^2\approx\phi^2-2\chi\phi$ with $\chi=\frac{\text{i} r}{1+r}$ and $\phi=\frac{\omega}{c}d$ \cite{vladimirova_exciton_1998}:
\begin{equation}
\tau(\omega_{10},\Omega_\text{c}) \approx \begin{cases}
\frac{(N-1)\Gamma_{10}(2\Omega_\text{c}^2-8\gamma_{20}^2)}{((4\gamma_{10} - 2\Gamma_{10})\gamma_{20} + \Omega_\text{c}^2)^2} &\phi\gg\chi\\
\frac{(N-1)\Gamma_{10}(2\Omega_\text{c}^2-8\gamma_{20}^2)}{((4\gamma_{10} - 2\Gamma_{10})\gamma_{20} + \Omega_\text{c}^2)^{3/2}}\frac{\sqrt{\phi}}{\sqrt{8\Gamma_{10}\gamma_{20}}} &\phi\ll\chi
\end{cases}
\label{delayanalytics}
\end{equation}
Figure~\ref{fig3} b shows both asymptotes of Eq.~\eqref{delayanalytics}, corresponding to the strong and weak Rabi-strength $\Omega_\text{c}$ regime ($\phi\gg\chi$, $\phi\ll\chi$). In the limit of $\gamma_{20}\rightarrow 0$, the condition $\phi\gg\chi$ is always fulfilled, and the corresponding $\tau(\omega_{10},\Omega_\text{c})$ is reduced to the formulas of textbook EIT \cite{fleischhauer_electromagnetically_2005}. In this case we find good agreement between the measured delays and Eq.~\eqref{delayanalytics}. In the opposite limit $\phi\ll\chi$, only qualitative agreement is observed. The systematic offset to the measured data is rooted in the band structure calculation of Eq.~\eqref{delayanalytics},
which assumes an infinite metamaterial an therefore deviates, in particular for small $\Omega_\text{c}$. More detail on the derivation of $\tau(\omega_{10},\Omega_\text{c})$ are provided in the Supplementary Material. Formally, EIT and ATS should be treated as two distinct, but closely related phenomena:  EIT is a destructive Fano-type interference effect between two excitation pathways, which can only occur if $\Omega_\text{c}<\xi \gamma_{10}$, smoothly transitioning to the ATS-regime for  $\Omega_\text{c}>\xi\gamma_{10}$, with two independent resonances of the dressed state doublet \cite{anisimov_objectively_2011}. The coefficient $\xi=(N^2-1)\phi/3$ is the approximate width of the bandgap for a finite system with $N<\pi/\phi\approx20$ qubits in units of $\gamma_{10}$. In this work, due to large $\gamma_{20}$, the accessible control strengths are in the transient regime $\Omega_\text{c}\gtrapprox \xi\gamma_{10}$. Here, a quantitative method to distinguish between ATS and EIT based on the Akaike's information criterion of Ref.~\cite{anisimov_objectively_2011} can not directly be applied to the measured data, since it features an asymmetric line shape due to the microwave background. Using the method for the calculated single-qubit splitting based on Eq.~\eqref{reflection} in conjunction with the experimentally extracted qubit decoherence rates yields that the observed line shape is by almost $100\,\%$ certainty described by the ATS (not shown).

We have verified the spectroscopically inferred time delays $\tau$ in pulsed time-domain measurements by using an FPGA-based heterodyne microwave setup (see Supplementary Material). Here, we generated Gaussian pulses of width $\sigma=50\,$ns at resonant center-frequency $\omega_{10}$ with a continuously applied control tone at frequency $\omega_{21}$ and strength $\Omega_\text{c}$. The pulse amplitude is kept at the single photon level ($P_\text{p}<\hbar\omega_{10}\Gamma_{10}$) in order to prevent saturation of the qubits. The digitized transmitted pulses for different control tone strengths are shown in Fig.~\ref{fig3} a. By fitting Gaussians to the measured data we extracted the temporal position of each pulse center. The extracted pulse arrival times were compared to a reference pulse, measured in the case of far detuned qubits, thus giving access to the pulse delay $\tau$. The delays measured in this way are presented in Fig.~\ref{fig3} b  showing good agreement with the spectroscopically inferred data and numerics. Due to the finite spectral width  $1/\sigma=20\,$MHz of the Gaussian pulses (indicated by dashed purple lines in Fig.~\ref{fig2} a) and that of the transparency window, the maximal accessible delay here is limited to $\tau\approx 12\,$ns ($n_\text{g}\approx 1500$).
In qualitative agreement with the transmission measurement data of Fig.~\ref{fig2}~a, the amplitude of the measured pulses becomes smaller when reducing the control tone amplitude. This effect is most prominent in the sector of large delays, however exactly here a high transmission is desired to realize an efficient quantum memory protocol \cite{novikova_electromagnetically_2012-1}. At maximal delay, the efficiency, i.e. the energy ratio of the transmitted and reference pulse, is about 16\,\%. Due to small average delay-bandwidth products of 0.2, the device under investigation appears suitable for a fixed-delay quantum memory \cite{rastogi_discerning_2019}, rather than a general purpose on-demand storage procedure, where the pulse has to fit spatially completely into the metamaterial for high storage efficiency  \cite{novikova_electromagnetically_2012-1}. Reducing the decoherence rate $\gamma_{20}$ would allow for significantly larger group indices and improve the bandwidth-delay product towards values larger than unity. Experimentally, this can be achieved by employing $\Lambda$-type three-level systems, such as flux- or fluxonium qubits \cite{leung_coherent_2012}.

\subsection{Dispersion-engineered slow light}
\begin{figure}[htb!]
	\includegraphics[width=\columnwidth]{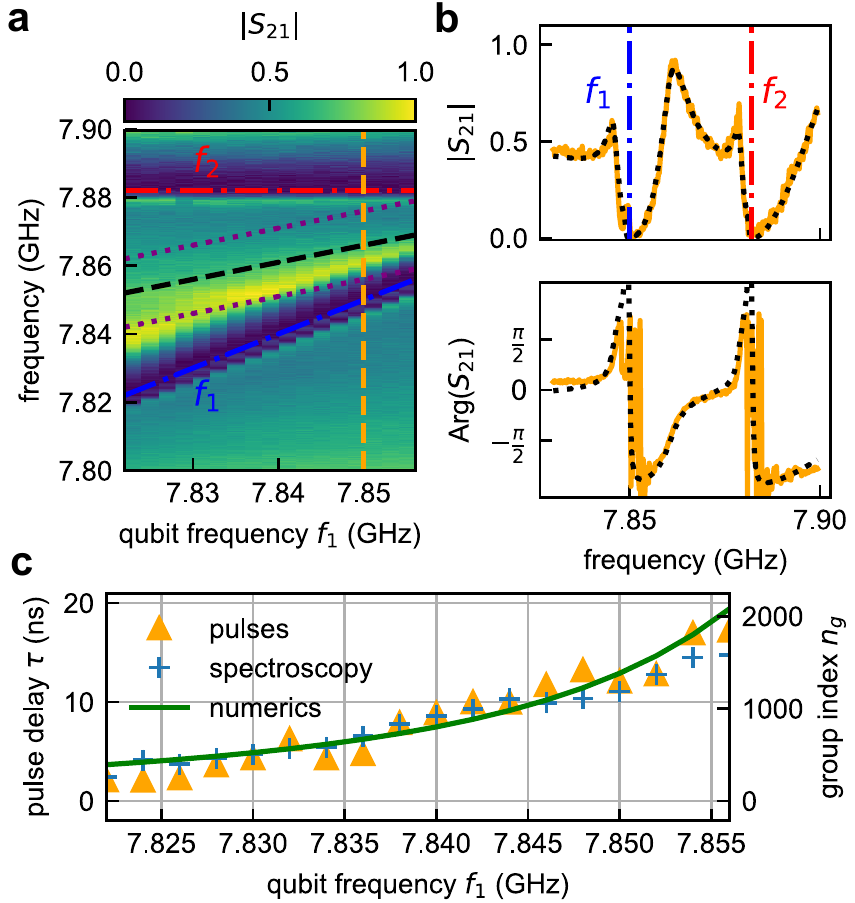}
	\caption{\textbf{a}. Absolute transmission $|S_{21}|$ for two collective four-qubit resonances at frequencies $f_1$ and $f_2$. The frequency is alternated between neighboring qubits along the waveguide. The black dashed line marks the center of the transparency window. The purple dashed lines indicate the bandwidth of the $50\,$ns Gaussian pulses. \textbf{b}. Transmission for a fixed detuning $f_2-f_1=32\,$MHz. The sharp peaks next to the bandgaps correspond to subradiant four-qubit states. The steep phase roll-off between the resonances indicates a low group velocity. Black dashed lines are fits to a numerical transfer matrix model. \textbf{c}. Experimentally extracted delays from pulsed and spectroscopic measurements in dependence on frequency $f_1$. Pulse delays up to $17\,$ns, corresponding to retardation factors of 1850, are achieved.}
	\label{fig4}
\end{figure}

As pointed out by Shen et al. \cite{Shen2007}, ATS-like band structures can be realized not only by dressed states, but also equivalently by two detuned distinct resonances coupled to the same mode continuum. For this approach, which we call \textit{dispersion engineering}, neither a third qubit level nor an additional microwave control tone is required. Here, we use the individual qubit frequency control to tune every second qubit to a frequency $f_1$ and every other qubit to $f_2$, creating two collective four-qubit resonances. The frequency $f_2=7.882\,$GHz is kept constant throughout the experiment, while the frequency $f_{1}$ of the other four qubits is varied. The transmission $S_{21}$ (Fig.~\ref{fig4} a, b) resembles that of the dressed state ATS in Fig.\ref{fig2}; the main ATS feature, being a transparency window with a steep phase roll-off between two bandgaps, is intact. Since the large $\gamma_{20}$ has no influence on the first two transmon levels, the brightest of the collective four qubit subradiant states are visible as pronounced peaks several MHz below $f_1$ and $f_2$ \cite{brehm_waveguide_2020} (compare Fig.~\ref{fig4} b). The observed line shape is in good agreement with a numerical transfer matrix calculation. The pulse delays $\tau$ and effective group indices $n_\text{g}$, calculated from Arg$(S_{21}(\omega))$ with Eq.~\eqref{finitesystemdelay} with respect to the frequency $f_{1}$, are shown in Fig.~\ref{fig4} c. Since the phase in the transparency window shows a varying slope, an average of the slope in a bandwidth of $10\,$MHz around the center is used to estimate $\tau$. Analogously to the previous section, we use Gaussian pulses with $\sigma=50\,$ns to directly measure and validate the expected pulse delay $\tau$. The center frequency is adjusted close to $(f_1+f_2)/2$ for each trace. Figure~\ref{fig4} c compares the delay $\tau$ of pulsed and spectroscopic measurements with numerical results from a transfer-matrix model. Here, we find delays up to $\tau=17\,$ns, corresponding to group indices of $n_\text{g}=1850$, which is comparable to the group indices found in the dressed-state ATS case for a similar splitting between the dressed states. In contrast to the ATS measurement, we achieve an efficiency of $\approx 45-50$\,\%, as the transmittance of the transparency window is solely limited by the non-radiative decoherence $\Gamma_\text{nr}$ and not influenced by $\gamma_{20}$. This also implies potentially possible much higher delays for narrower transparency windows. Even though the dispersion-engineered approach lacks the possibility of fast and simple control via an additional microwave tone in contrast to the ATS, it can be used as a fixed delay quantum memory and showcases the ability to tailor the band structure on demand in superconducting wQED. For further studies, one can think of more complex band structures going beyond the capabilities of the ATS, e.g., engineering several spectral regions of different group indices, referred to as multi-color slow light \cite{Wang2014}.

\section{Summary \& Conclusion}
In this work, we demonstrated slow light in a superconducting waveguide QED system. A metamaterial consisting of eight densely spaced transmon qubits was used to engineer a band structure with a flat dispersion profile to obtain reduced group velocities of electromagnetic waves. Using the well-known dressed-state ATS, we observed group velocities reduced by a factor of down to 1500, both in spectroscopic and direct time-resolved pulsed measurements. The maximum achievable delay and efficiency is limited by the strong dephasing of the $2\rightarrow 0$ transition, which is inherent to ladder-type three-level systems. Moreover, we demonstrated slow light with a tailored band structure based on distinct detuned collective resonances of the qubits. At comparable retardation, we observed three times larger efficiencies, not limited by large $\gamma_{20}$. The demonstrated slow-light metamaterial can be employed as a fixed-delay quantum memory or a tunable pulse retarder to synchronize pulses in quantum information processors. Based on our proof-of principle experiment, further improvements which significantly ameliorate the device performance can be made. A similar metamaterial based on flux or fluxonium qubits, which feature a $\Lambda$-type level structure, is expected to reach the EIT regime and may ultimately realize a general purpose and on-demand superconducting quantum memory.

\section{Acknowledgements}
This work has received funding from the Deutsche Forschungsgemeinschaft (DFG) by the Grant No. US 18/15-1, the European Union’s Horizon 2020 Research and Innovation Programme under Grant Agreement No. 863313 (SUPERGALAX), BMBF Grant 'PtQUBE' No. 13N15016, and by the Initiative and Networking Fund of the Helmholtz Association, within the Helmholtz Future Project ‘Scalable solid state quantum computing’. JDB acknowledges financial support from Studienstiftung des Deutschen Volkes, AS was supported by Landesgraduiertenf\"orderung-Karlsruhe and ANP gratefully acknowledges support from Russian Science Foundation Grant 20-12-00194. AVU acknowledges partial support from the Ministry of Education and Science of the Russian Federation in the framework of the Program of Strategic Academic Leadership ”Priority 2030”.

\clearpage
\bibliography{SlowLight_bib}
\onecolumngrid

\newpage

\appendix

\setcounter{figure}{0}
\setcounter{equation}{0}
\renewcommand{\thefigure}{S\arabic{figure}}
\renewcommand{\thetable}{S\arabic{table}}
\renewcommand{\theequation}{S\arabic{equation}}

\section{Qubit characterization}
\label{qubitcharct}
Extensive details on the characterization of the individual qubits can be found in the Supplementary Material of reference \cite{brehm_waveguide_2020}. The individual qubit characteristics around 7.81 GHz are listed in table \ref{singlequbitparamsgeneral}.

\begin{table}
	\caption{Measured\footnote{Measured after thermal cycle of the cryostat compared to the measurements in the main text.} individual qubit properties around $7.81\,$GHz and upper- and lower-sweet spot positions $f^\text{min}_{01}$, $f^\text{max}_{01}$.}
	\begin{ruledtabular}
		\begin{tabular}{c| c c c c c c c c}
			Parameter & Qubit 1 & Qubit 2 & Qubit 3 & Qubit 4 & Qubit 5 & Qubit 6 & Qubit 7 & Qubit 8\\
			$\Gamma_{10}/2\pi\,$(MHz)&7.3&  9.5&  11.3&  13.9&
			14.5& 14.6& 12.1&  11.9\\
			
			$\gamma_{10}/2\pi\,$(MHz)&4.2&  5.3&  6.7& 8.4&
			8.1&  8.1&  6.7& 7.2\\
			
			$\Gamma_\text{nr}/2\pi\,$(MHz)&0.52&  0.56& 1.0&  1.36&
			0.83&  0.83&  0.65&  1.3 \\
			$\gamma_{20}/2\pi\,$(MHz)\footnote{$\gamma_{20}$ was not accessible for all qubits at the given frequency, due to spurious TLS distorting the single qubit ATS.}&8.7&   8.3&   -&   6.7&
			5.6&  6.2& -&  5.7 \\
			
			Ext. coeff.$\,\%$&98.4&  98.9&  97.7&  97.4&
			99.0& 99.0& 99.1 & 96.8\\

			$\chi/2\pi\,(\text{MHz})$\footnote{Measured at 7.9 GHz}&283&279&273&275&267&281&273&276\\
			$f^\text{max}_{01}\,(\text{GHz})$&8.097&7.900&8.088&8.114&8.115&7.95&8.066&8.136\\ 
			$f^\text{min}_{01}\,(\text{GHz})$&3.029&3.091&2.912&2.986&2.970&2.936&2.588&2.484\\
		\end{tabular}
	\end{ruledtabular}
	\label{singlequbitparamsgeneral}
\end{table}

\section{Experimental microwave setup}
\begin{figure}[htb!]
	\includegraphics[width=\textwidth]{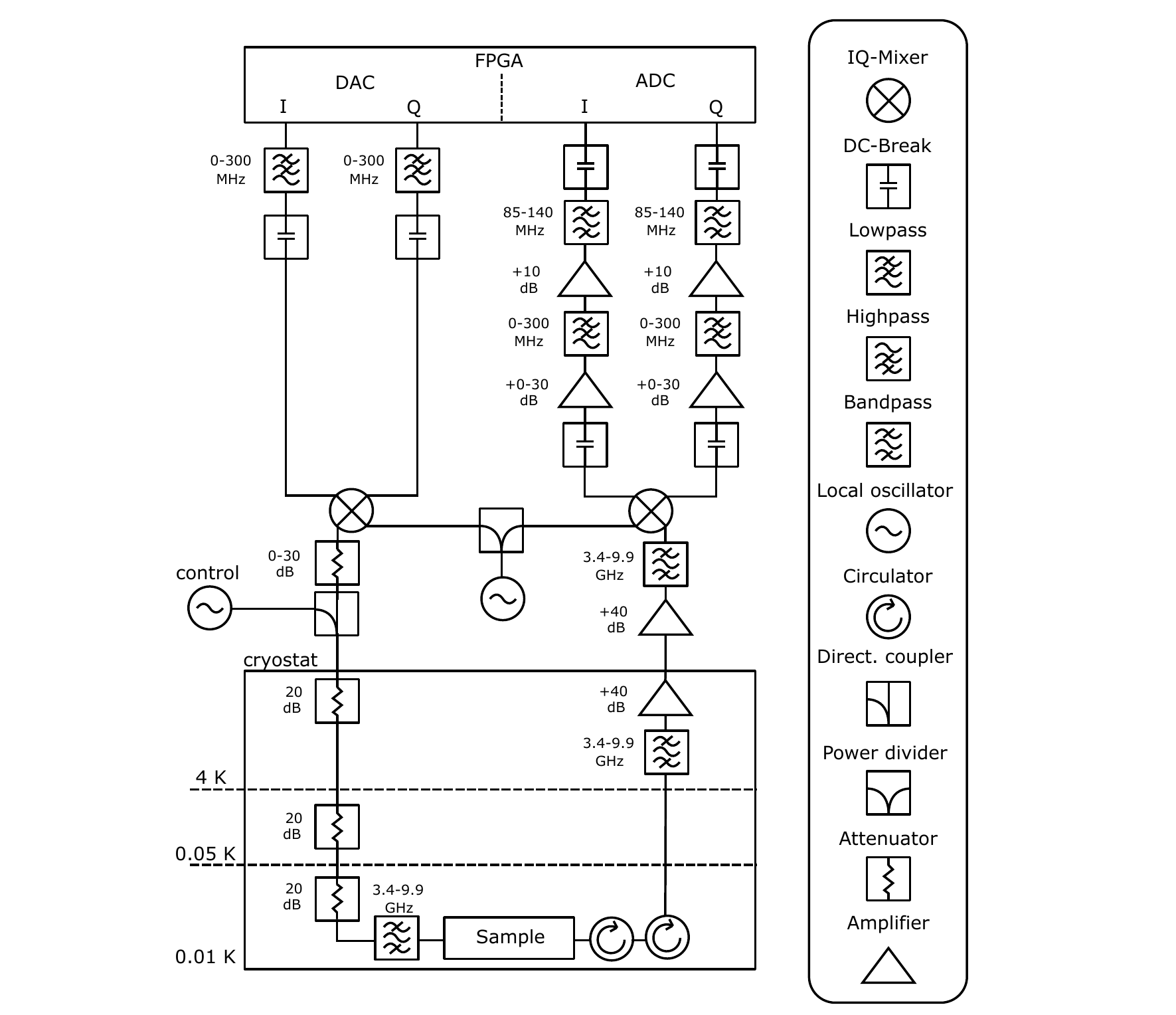}
	\caption{Used microwave setup for time-resolved measurements with heterodyne detection of microwave pulses.}
	\label{setup}
\end{figure}

The used microwave setup for the time-resolved measurements is shown in Fig.~\ref{setup}. It is controlled by a Xilinx ZCU111 evaluation board featuring an RFSoC architecture. This combines CPUs, an FPGA, DACs and ADCs on a single chip. The setup is operated with custom firmware and an effective sampling rate of 1 GS/s. The microwave pulses are generated and detected at an intermediate frequency of $f_\text{IF}=115\,$MHz. Single-sideband mixing with a microwave local oscillator based on IQ-mixing is employed to up- and down-convert the signal between $f_\text{IF}$ and Gigahertz frequencies. After detection, the measured pulses are digitally downconverted from the intermediate frequency to dc and low-pass filtered with a 5th order Butterworth filter with a $115\,$MHz cutoff.

\section{Calibration of absolute power with the ATS and extraction of $\gamma_{20}$}
\begin{figure}[htb!]
	\includegraphics[width=\textwidth]{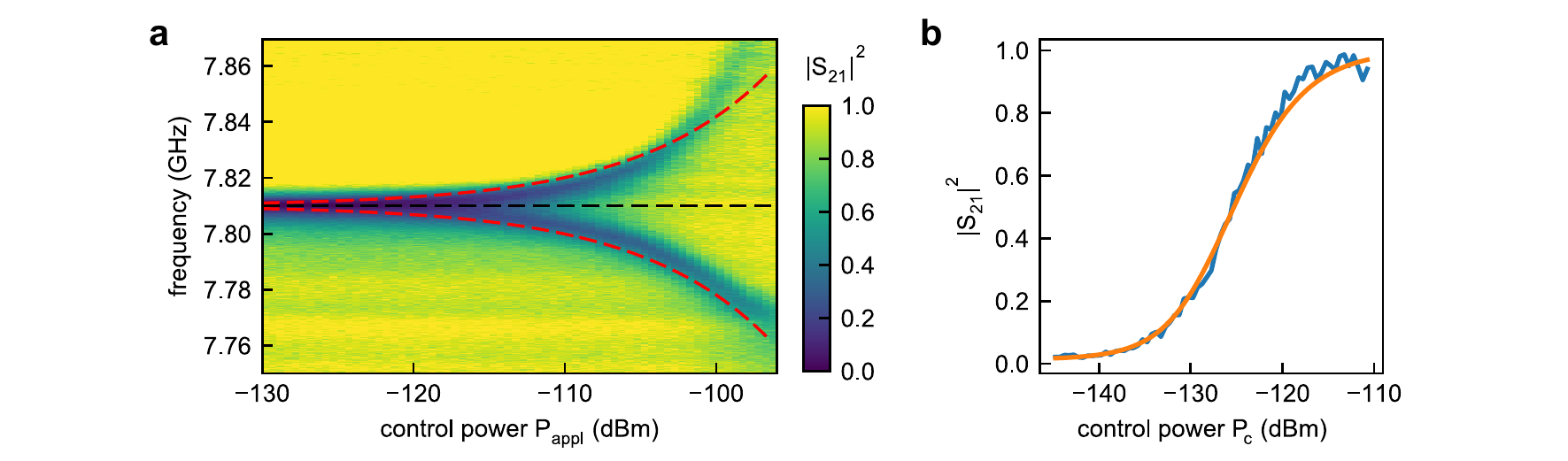}
	\caption{\textbf{a}. Autler-Townes splitting of qubit 2 as a function of the applied power $P_\text{appl}$ used to calibrate the absolute on-chip power $P_\text{c}$. Red dashed line is a fit to $\Omega=a\sqrt{P_\text{appl}}$. \textbf{b}. Resonant transmission ($\omega=\omega_{10}$) as function of $P_\text{c}$ used to extract $\gamma_{20}$. The orange line is a fit to \ref{3Lonres}. For both figures $|S_{21}(\omega_{10})|$ is normalized to 1 for large $\Omega_\text{c}$.}
	\label{powercalib}
\end{figure}
The Rabi strength $\Omega_{j+1j}$ of the coupling of the $j+1\rightarrow j$ transition of a transmon to a single mode is given by \cite{koch_charge-insensitive_2007,hoi_microwave_2013}:
\begin{equation}
\hbar \Omega_{j+1j}=2e\frac{C_\text{c}}{C_\Sigma}\braket{j+1|\hat{n}|j} V\approx 2e\frac{C_\text{c}}{C_\Sigma}\sqrt{j+1}\left(\frac{E_\text{J}}{8E_\text{c}}\right)^{1/4} V_\text{rms}=\mu_{j+1j} V_\text{rms}
\end{equation}
$\mu_{j+1j}$ is denoting the dipole moment of the corresponding transition and $V_\text{rms}$ is the root-mean square voltage of the incoming wave ($P_\text{c}=\frac{V_\text{rms}^2}{Z_0}=\frac{V_\text{c}^2}{2Z_0}$, $Z_0=50\,\Omega$ is the impedance of the waveguide). 
The rate of incoming photons to the sample is given by:
\begin{equation}
\nu=\frac{P_\text{c}}{\hbar \omega_{j+1j}}=\frac{V_\text{rms}^2}{Z_0\hbar \omega_{j+1j}}=\frac{\hbar\Omega_{j+1j}^2}{Z_0\omega_{j+1j}\mu_{j+1j}^2}
\end{equation}
With the relaxation rate $\Gamma_{10}=\frac{\omega_{10}^2C_c^2Z_0}{2C_\Sigma}$ this can be written as:
\begin{equation}
\nu=\frac{\Omega_{j+1j}^2}{2(j+1)\Gamma_{10}}
\end{equation}
In case of the ATS, the $2\rightarrow 1$ transition is driven with a control tone. For this specific case we get the following connection between $\Omega_c=\Omega_{21}$ and the incident power:
\begin{equation}
P_\text{c}=\frac{\Omega_\text{c}^2}{4\Gamma_{10}}\hbar\omega_\text{c}
\end{equation}
The measured single-qubit ATS, as shown in Fig.~\ref{3Lonres} is fitted with a simple $\Omega=a\sqrt{P_\text{appl}}$ law. The calibration factor $\alpha$ between the applied power $P_\text{appl}$ and the correct incident on chip-power $P_\text{c}$ ( $P_\text{c}=\alpha  P_\text{appl}$) is calculated with:
\begin{equation}
\alpha=a^2\frac{\hbar \omega_\text{c}}{4\Gamma_{10}}
\end{equation}
We note that the calibration depends on the radiative relaxation rate of the qubits.\\

Once the power is calibrated, we extract the decoherence rate $\gamma_{20}$ of the $2\rightarrow 0$ transition. The complex reflection coefficient of a dressed 3-level system, as derived in reference \cite{abdumalikov_electromagnetically_2010}, is given by:
\begin{equation}
r=-\frac{\Gamma_{10}}{2[\gamma_{10}-\text{i}(\omega-\omega_{10})]+\frac{\Omega_c^2}{2\gamma_{20}-2\text{i}(\omega-\omega_{10}+\omega_c-\omega_{21})}}
\label{3L}
\end{equation}
$\Omega_\text{c}$ is the Rabi-strength of the control tone with frequency $\omega_\text{c}$. If the probe tone is resonant with the $0\rightarrow 1$ transition ($\omega=\omega_{10}$) and the control tone is resonant with the $1\rightarrow 2$ transition ($\omega_\text{c}=\omega_{21}$) the corresponding transmission coefficient of Eq.~\eqref{3L} simplifies to:
\begin{equation}
t=1-\frac{\frac{\Gamma_{10}}{2\gamma_{10}}}{1+\frac{\Omega_c^2}{4\gamma_{20}\gamma_{10}}}=1-\frac{\frac{\Gamma_{10}}{2\gamma_{10}}}{1+\frac{4\Gamma_{10}}{4\gamma_{20}\gamma_{10}\hbar\omega_\text{c}}P_\text{c}}
\label{3Lonres}
\end{equation}
When $\Gamma_{10}$ and $\gamma_{10}$ are known (see table \ref{singlequbitparamsgeneral}), a fit of Eq.~\eqref{3Lonres} to the measured trace gives an estimate for $\gamma_{20}$, compare figure \ref{powercalib} b).\\

The rates $\Gamma_{10}$, $\gamma_{10}$, and $\gamma_{20}$ used for the numerical calculation in Fig.~\ref{fig3} of the main text are averaged values of all individual qubit rates. We note that the measured decoherence rates are frequency dependent, because of the random distribution of parasitic two-level-systems over the qubits' flux dispersion, the varying influence of magnetic flux noise, and the influence of standing waves in the microwave background of the cryostat. Due to this frequency dependence, the decoherence rates are left as free fitting parameters for the numerical calculation in Fig.~\ref{fig4} to get a more accurate estimate for the effective rates.
\section{Band structure calculation and derivation of $\tau(\Omega_c)$ in the metamaterial limit}
The band structure of an infinitely extended  metamaterial, composed of three-level systems under ATS condition, as given by Eq.~\eqref{3L}, can be calculated as an eigenvalue problem of the corresponding T-matrices:
\begin{equation}
T_1T_\phi
\begin{pmatrix}
V_1^R\\V_1^L
\end{pmatrix}=\exp(\pm\text{i}kd)\begin{pmatrix}
V_1^R\\V_1^L
\end{pmatrix}
\label{blochtheorem}
\end{equation}
$V_1^{R/L}$ denote right and left propagating fields. $T_1$ is the transfer matrix of a single qubit and $T_\phi$ of a bare piece of transmission line leading to a phase shift of $\phi=\omega/cd$. More details on the transfer matrix formalism are provided in the Supplementary Material of Ref.~\cite{brehm_waveguide_2020}. The eigenvalue problem of Eq.~\eqref{blochtheorem} reduces to:
\begin{equation}
\cos(kd)=\frac{1}{2}\Tr(T_1T_\phi)=\cos\left(\frac{\omega}{c}d\right)+\frac{\text{i} r}{1+r}\sin\left(\frac{\omega}{c}d\right)
\label{bandfreqs}
\end{equation}
Solving Eq.~\eqref{bandfreqs} for $\omega(k)$ gives access to the dispersion relation and band structure.
In the large wavelength-limit, meaning $kd\ll1$, $\omega/cd\ll1$ this equation can be approximated as:
\begin{equation}
1-\frac{1}{2}(kd)^2\approx 1-\frac{1}{2}(\frac{\omega}{c}d)^2+\frac{\text{i} r}{1+r}\frac{\omega}{c}d
\end{equation}
\begin{equation}
(kd)^2\approx\phi^2-2\chi\phi\quad\text{with}\quad\chi=\frac{\text{i} r}{1+r},\,\phi=\frac{\omega}{c}d
\label{dispeqapprox}
\end{equation}
In order to obtain an analytical expression for $v_\text{g}=\text{Re}(\left(\frac{d k }{d\omega}\right)^{-1})$, Eq.~\eqref{dispeqapprox} can be approximated in the limits of large and small control tone strengths $\Omega_\text{c}$, translating to $\phi\gg\chi$ and $\phi\ll \chi$.
For the first case, Eq.~\eqref{dispeqapprox} simplifies to:
\begin{equation}
k\approx \frac{\omega}{c}-\frac{\text{i} r}{1+r}\frac{1}{d}
\end{equation}
The inverse group velocity:
\begin{equation}
\frac{1}{v_\text{g}}=\frac{dk}{d\omega}\approx \text{Re}\left(\frac{1}{c}-\frac{1}{d}\frac{d}{d\omega}\left(\frac{\text{i} r}{1+r}\right)\right)
\end{equation}
at $\omega=\omega_{10}$ this equates to:
\begin{equation}
\left.\frac{1}{v_\text{g}}\right\vert_{\omega=\omega_{10}}=\frac{1}{c}+\frac{1}{d}\frac{\Gamma_{10}(2\Omega_\text{c}^2-8\gamma_{20}^2)}{((4\gamma_{10} - 2\Gamma_{10})\gamma_{20} + \Omega_\text{c}^2)^2}
\label{groupvel}
\end{equation}
The pulse delay of an array of $N$ qubits with a spacing $d$ is then given by:
\begin{equation}
\tau=\frac{(N-1)d}{v_\text{g}}
\end{equation}
We note that for a small decoherence rate of the $2\rightarrow 0$ transition, $\gamma_{20}\approx 0$, the group velocity of textbook EIT can be recovered from Eq.~\eqref{groupvel}:
\begin{equation}
\frac{1}{v_\text{g}}=\frac{1}{c}+\frac{1}{d}\frac{2\Gamma_{10}}{\Omega_c^2}
\end{equation}
For the second case, i.e. $\phi\ll\chi$, Eq. \ref{dispeqapprox} simplifies to:
\begin{equation}
k^2\approx -\frac{2\phi\chi}{d^2}
\end{equation}
\begin{equation}
\frac{dk}{d\omega}\approx\frac{d\chi}{d\omega}\frac{\sqrt{-\phi/2}}{\sqrt{\chi}d}
\end{equation}
Using the principal square root, the inverse group velocity at $\omega=\omega_{10}$ is then given by:
\begin{equation}
\frac{1}{v_\text{g}}=\text{Re}\left(\left.\frac{d k }{d\omega}\right\vert_{\omega=\omega_{10}}\right)=\frac{1}{d}\frac{\Gamma_{10}(2\Omega_c^2-8\gamma_{20}^2)}{((4\gamma_{10} - 2\Gamma_{10})\gamma_{20} + \Omega_c^2)^{3/2}}\frac{\sqrt{\phi}}{\sqrt{8\Gamma_{10}\gamma_{20}}}
\end{equation}
\subsection{Normalization of spectroscopic data}
All spectroscopic data are normalized by  $S_{21}(\omega)=S^\text{meas}_{21}(\omega)/(aS^\text{bg}_{21}(\omega))$. Here, $S^\text{bg}_{21}(\omega)$ is the measured transmission coefficient of the background which is obtained when all qubits are far detuned. $a$ is a constant factor accounting for interference effects of the signal with the background and fluctuations of the amplifier gain.
\end{document}